# Quantitative and qualitative analysis of editor behavior through potential coercive citations

Claudiu Herteliu[1,*] Marcel Ausloos[2,3,*] Bogdan Vasile Ileanu[1]
Giulia Rotundo[4] Tudorel Andrei[1]

[1]Department of Statistics and Econometrics, Bucharest University of Economic Studies, Bucharest, Romania, [2]School of Business, University of Leicester, Leicester, UK [3]GRAPES – Group of Researchers for Applications of Physics in Economy and Sociology, Liege, Belgium, [4]Department of Methods and Models for Economics, Territory, and Finance, Sapienza University of Rome, Rome, Italy

[*]Correspondence address. E-mail: hertz@csie.ase.ro, claudiu.herteliu@gmail.com and marcel.ausloos@ulg.ac.be, ma683@le.ac.uk

June 7th, 2017

## Abstract

How much is the h-index of an editor of a well ranked journal improved due to citations which occur after his/her appointment? Scientific recognition within academia is widely measured nowadays by the number of citations or h-index. Our dataset is based on a sample of four editors from a well ranked journal (impact factor – IF – greater than 2). The target group consists of two editors who seem to benefit by their position through an increased citation number (and subsequently h-index) within journal. The total amount of citations for the target group is bigger than 600. The control group is formed by another set of two editors from the same journal whose relations between their positions and their citation records remain neutral. The total amount of citations for the control group is more than 1200. The timespan for which citations' pattern has been studied is 1975-2015. Previous coercive citations for a journal benefit (increase its IF) has been signaled. To the best of our knowledge, this is a pioneering work on coercive citations for personal (editors') benefit. Editorial teams should be aware about this type of potentially unethical behavior and act accordingly.

**Keywords:** academic journal editor; citations; inappropriate tactics; coercive citations; citations overdose

## Introduction

In an era when information spreads almost instantly and interconnections between academic networks play a crucial role, it is very clear that prestigious journals represent an important stakeholder for knowledge and its advancement [1-2]. Therefore, to be the editor of an academic journal, despite of doing it *pro bono*, represents (or at least should represent) one of the most important positions for influencing one's own community, particularly when it comes to the most contemporary research topics. Scholars in this position should act with integrity, responsibility (for the journal's audience and contributors), and neutrality [3], restraining his or her passion for such topics. Yet editors, we must admit, do have excuses: like other academics, they struggle in order to be hired, promoted, gain grants, publish in top journals, and enhance their recognition within academia, in addition to the routine labor of finding sufficient reviewers.

On the other side, that of authors, almost every person involved in academic life, from graduate students and up to tenured professors, wants to publish their ideas and research outcomes in journals ranked as highly as possible. Hence, during a peer review process, difficult situations that can lead to unfair practices may occur [4]. Incidents of misconduct have been reported that involve data fabrication, faking peer review, duplication, reproducibility, and plagiarism on the part of authors [5-8]. Editors themselves can undermine the peer review process in much the same way that authors or third-party agencies have done elsewhere [9].

On the editors and reviewers' side, unethical or immoral incidents may also occur [10-14]. Perhaps one of the most famous recent cases of academic editorial misconduct has concerned *Chaos, Solitons, and Fractals* (CSF) ex-editor in chief, Mohamed El Naschie, who publish five papers of his own in an issue of 36 articles in volume 38 (5), 2008, of CSF [10, 15].

Ethical standards state that editors and reviewers should refrain from the "practice of coercive citations" [16]. For the current study a coercive citation is considered to occur when an editor requires that the manuscript's author(s) include unnecessarily, "moderately relevant" (or "rather unessential for the submitted paper's understanding") reference(s) to the editor's publication(s) [11, 17].

Usually, such "coercive citations" are suggested in order to increase the reputation of some research entity or group leader, but they can also improve the impact factor of a specific journal and hence the "market share" of the editors or the publisher. Nevertheless, self- or network citations are not an entirely undesirable situation *per se*: sometimes such situations may highlight new research topics or "scientists' field mobility" and hence creativity [18].

Pursuing our consideration of the quantity of citations of research work, "coercive citations" required by peer reviewers are hard to prove. Data on the peer review process is very difficult to obtain from editors (mostly because it is confidential), while authors rarely communicate reviews (though authors do acknowledge [anonymous] reviewers inputs). However, it is very easy to observe citations of the works of journal editors.

Network citations or self-citations are not so uncommon [17, 19]. Their role has been discussed in many papers, often in order to criticize or refine the *h*-index [20-21]. "*A scientist has index h if h of his or her $N_p$ papers have at least h citations each and the other ($N_p$ - h) papers have ≤h citations each*" where $N_p$ is number of published papers [22]. However, many academics attempt to enhance their own *h*-index [23-25]. This is especially the case in disciplines where it is known or required that the *h*-index is one of the hiring or promotion criteria [26-27]. Concerns with and deficiencies of this system are sometimes pointed out by those who object to the *h*-index or to how it is calculated [28]. In our opinion, it is a quality criterion, to be considered like many other subjective measures. Significantly, since the evolution of an editor *h*-index can be easily observed, it might be possible to obtain a correlation with any particular editor's presence on or addition to an editorial board.

In fact, the current study provides a case study of authors' paper citation accelerations and the resulting *h*-index increase. It will be shown that citations of works published by editors (and co-workers), and subsequently their *h*-index, can be and is necessarily increased by the editorial filtering methods. Furthermore, the current paper provides several case studies of editors who have had their level of recognition significantly increased after joining an editorial board. Indeed, the editors might have accidentally improved their h-index on account of being concerned with accuracy in quoting published work. A collateral positive effect, the recognition increase of a co-author, will be shown as well.

Peripherally, in this study we will be concerned about the role of special issues. There is of course nothing wrong in citing an editor (of a special issue or proceedings of a conference) for his/her particular achievement, in particular in papers published through such means. In fact, it has been shown that there is some fair balance between co-authorship and scientific recognition indices for classical peer review papers and "special issues" (in a broad sense, i.e., including proceedings outside of journals) [29-31]. Thus, it is good for the scientific peer review process that new editors appear on editorial boards, especially when an individual has demonstrated editorial skill. However, it is also of interest, within our peer review investigation framework, to focus on how and where the reputation of an editor grows.

## Materials and methods

The core of the present report deals with papers published by three authors (Author1, Author2, and Author3) in Journal1. Author1 and Author2 belong to the editorial board of Journal1. In order to outline the detected effect, i.e. how editors' *h*-index increases, we first outline the complete publication list of such authors in various time intervals and the citations or lack of citations of such papers in Journal1, taking into consideration various time intervals. Two additional editors from Journal1 (Author 4 and Author 5) have been added as a control sample.

Before starting with the presentation of the material and methods several key terms that will be used must be clarified.

Coercive citation: Wilhite [17] suggest that requesting self citations does *"(i) give no indication that the manuscript was lacking in attribution; (ii) make no suggestion as to specific articles, authors, or a body of work requiring review; and (iii) only guide authors to add citations from the editor's journal"* represents a coercive citation situation on the part of the editorial staff. An adaptation of this definition for our current purpose should remove item (ii) and replace "journal" from item (iii) with "papers." Without labeling it "coercive citation", Resnik [11] described such an event as when a *"reviewer required you to include unnecessary references to his/her publication(s)."* Setting out from these prior definitions, the current report considers that a coercive citation occurs when an editor requires that the manuscript's author(s) include "moderately relevant" (or "rather unessential for the submitted paper's understanding") reference(s) to his/ her publication(s).

Self-citation: Let X be author of paper H. When X writes a new paper (K) and cites paper H, this event is called a self-citation for X.

Semi-self-citation: Let X and Y be co-authors of paper H. When Y writes a new paper (K) (without having X as co-author) and cites paper H, this event is called a semi-self-citation for X.

H-index: The *h*-index "defines" the core of papers of an author from the relationship between the number of citations $n_c$ and the corresponding rank $r$ of a paper, through a simple threshold $r_c$, i.e. if $n_c \geq r_c$, then $r_c = h$ [22, 32].

Special Issue: Approaches and procedures for special issues vary across journals and disciplines. Sometimes these issues are based on selected papers from academic conferences. Within the current report, it appears that special issues published by Journal1 were not based on papers from scientific conferences. Even so, the peer review process for items included within Journal1's special issues seems to be different from their standard review process (for manuscripts submitted for publication in a regular issue). There are two points that this statement relies on. First, it is based on the history section for each paper. In the case of a regular paper there are steps that are usually well known: initial reception, revised version (if any), acceptance, online availability. For papers belonging to Journal1's special issues, only one date is mentioned: online availability. The second point is based on personal communication from the Editor in Chief (EIC) of Journal1. He mentioned that Author1 *"was a quest editor for a special issue. A quest editor is responsible for the peer review process."* (N.B. The original grammar is kept, here and in all quotes.)

The current manuscript's approach in finding potential coercive citations is presented in figure 1.

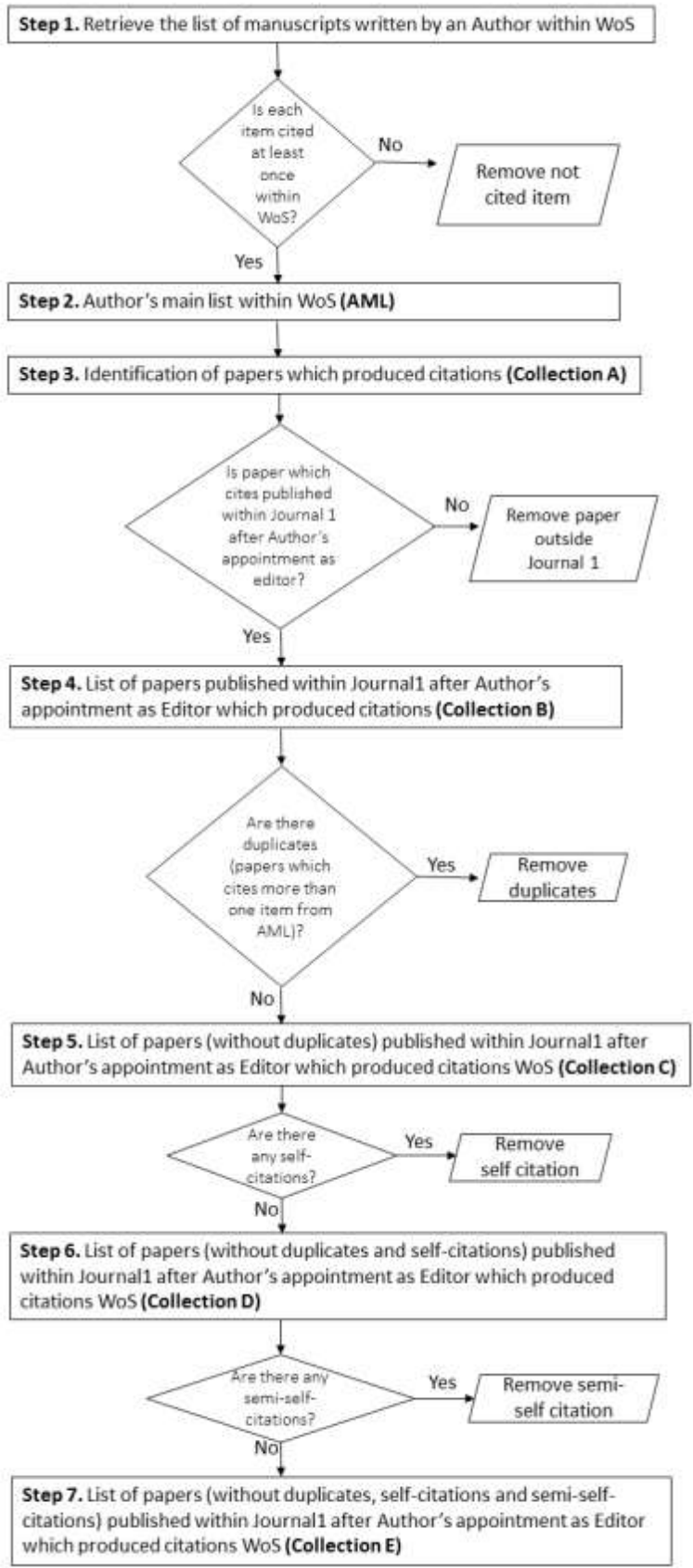

**Figure 1.** Methodological framework

Nowadays the academic community is using three major multi-disciplinary scientific databases (Web of Science Thomson Reuters (WoS), Scopus, and Google Scholar) in an extensive way. It is admitted that there is no ideal database [33]. Each has strengths and weaknesses. For the current research paper, WoS has been chosen as the data source.

A synopsis of information with regards to our current research framework is presented in table 1.

**Table 1.** WoS's publications and citations records for Author1, Author2, Author3, Author4, and Author5 as were registered after performing our research framework

| Research's step | Author1 | Author2 | Author3 | Author4 | Author5 |
|---|---|---|---|---|---|
| Retrieval time | Nov. 18, 2015 | From Nov. 25, 2015 to Dec. 3, 2015 | Dec. 22, 2015 | Dec. 2, 2016 and Jan. 4, 2017 | Dec. 2, 2016 and Jan. 4, 2017 |
| **Step 1** items found in WoS | 46 | 55 | 44 | 196 | 47 |
| from which within Journal1 | 10 | 15 | 2 | 4 | 3 |
| **Step 2** author's main list (AML) | 34 | 32 | 11 | 77 | 39 |
| **Step 3** total received citations (Collection A) | 310 | 303 | 52 | 958 | 345 |
| from which within Journal1 | 130 | 183 | 29 | 48 | 7 |
| **Step 4** Collection A from Journal1 after editorial appointment (Collection B) | 123 | | | | |
| **Step 5** Collection B without duplicates (Collection C) | 69 | | | | |
| **Step 6** Collection C without self-citations (Collection D) | 67 | | | | |
| **Step 7** Collection D without semi-self-citations (Collection E) | 65 | | | | |

**Note:** In order to have the same timespan for the all analyzed authors, the post 2015 publications and citations for Author4 and Author5 have been ignored.

The list from A1_Collection_E was sent to the Journal1 editor-in-chief, in order to confirm (or not) that the responsible editor of these papers was Author1. Their answer stated that Author1 was editor for a special issue and, subsequently, was responsible for the peer review process.

In parallel, all (65) corresponding authors of papers from A1_Collection_E were approached with two questions: (Q1) to confirm whether or not Author1 was the responsible editor of their paper and (Q2) to confirm whether or not, during peer review process, suggestions for citing Author1's paper(s) occurred.

We knew from the reference lists within their manuscripts which paper(s) of Author1 they cited. Therefore, the second question (Q2) was a closed one by providing a specific list of paper(s). Most of the time (table 2), the list contains only one item but once there were six items on the list.

**Table 2.** Number of Author1's manuscripts enlisted within the second question (Q2)

| Manuscripts enlisted | Cases | Expected citation overdose |
|---|---|---|
| 1 | 40 | No |
| 2 | 14 | |
| 3 | 6 | |
| 4 | 2 | |
| 5 | 2 | |
| 6 | 1 | Yes |
| Total | 65 | |

**Note:** Information from the right hand side column is according to our own subjective perception.

A small proportion (nine persons) provided an answer to the questions posed. Various answers were received for both items. Regarding the first question, the response was usually "no" or "don't know." For the second question, when the answer to the first question was "yes" (three cases out of nine responses) "coercive citation" of Author1's papers occurred twice. When the answer to the first question was "no" or "don't know" interesting information appears regarding a second Journal1's editor (Author2 as may be seen within the sub-section from the Results section) (all journal's names listed here (underlined) and scientific subdomain were anonymized by the authors of this article):

> *"i was asked to reference this list of semi relevant manuscripts"* or
>
> *"in the review process, (revised form) recommended to me many articles by one of the referees (about six). I examined them carefully and only two of thems were added references list. one of them is your written manuscript."* or
>
> *"I was asked to revised my article twice. The first time I did not get any suggestion to enhance the reference list. The second time I got the following:*

> *-Bibliography on <u>scientific subdomain</u> must be updated. Useful references can be found in recent publications in <u>Journal1</u>, <u>Journal2</u>, <u>Journal3</u>, <u>Journal4</u>, among others. Additional references could be: Here three references where mentioned. Among which the one you asked me about was suggested."*

This approach for first signaling potential journals where coercive citations occur is in line with a very recent paper [41]. The paper started from the list provided by Wilhite [17] and selected for further bibliometric analysis all journals which record at least two observations of perceived coercive citations, without any specific justification of this threshold= 2. In the following we stress (step1 to step7) the case of the Author1 because he/she has the largest h-index for citations provided by papers published outside Journal1. For all other authors we stop after step3 for conciseness (table 1).

In our case, the above mentioned answers clearly show that coercive (according to our definition) citations occur. Indeed, we are not sure about who suggested the addition of those titles. It could be one (or more) external reviewer(s) or the editor. Let us hypothetically suppose that there is an external reviewer who suggests that six papers (co-)authored by the editor be added to the references list. Of course, that depends on the manuscript's topic being under review and subsequently on suggested references fit to that topic. From our point of view, this hypothetical case can be considered as a "<u>citations overdose</u>". A neutral editor could not simply concatenate external reviewers' reports. In case of too many suggestions about his/ her (editorial) work, a selection can be made by the editor.

The statistical association was Chi Square tested. The maximum significance level was set to be equal to 0.05. In line with other research [45], we distinguish the impact of editing of special issue versus a regular issue through a Chi Square test.

**Results**

If the study is run using the three aforementioned data bases (WoS, Science Direct, and Google Scholar), different results are achieved due to variations in the details that each contains. However, the overall conclusions do not significantly differ [33]. Thus we are reporting only the data from WoS. Moreover, in order to be as general as possible, we will use the word "item" to facilitate inclusion of what are not strictly "papers in peer review journals."

According to WoS [34] and based on its Impact Factor (IF) ranking, the Journal1 IF is greater than 2 in 2015, 2014, 2013, 2012, respectively; Journal1 is considered to be a top journal within the Academic1 domain. Indeed, WoS Journal Citation Reports (JCR) 2015 ranked Journal1 on the 8[th] decile within the "Academic1_1" subdomain and on the 7[th] decile within the "Academic1_2" subdomain.

1. **Author1**

On January 2013 (Journal1's first issue in that year) a new member of the editorial team was appointed: Author1. From WoS, Author1 has (co-)authored 46 papers (33 prior to 2013). He/she published 10 papers with Journal1 (only three prior to 2013). His/her h-index is 10 in 2015 compared to seven in 2012 (Table 3).

**Table 3.** Time dependence of the *h*-index for Author1, Author2, Author3, Author4, and Author5, distinguishing between citations of Author1, Author2, Author3, Author4, and Author5 papers, wherever published, but cited or not in Journal1

| *h*-Index | 1996 | 1997 | 1998 | 1999 | 2000 | 2001 | 2002 | 2003 | 2004 | 2005 | 2006 | 2007 | 2008 | 2009 | 2010 | 2011 | 2012 | 2013 | 2014 | 2015 |
|---|---|---|---|---|---|---|---|---|---|---|---|---|---|---|---|---|---|---|---|---|
| Author1-all | | | | | | | | | | | 1 | 2 | 4 | 4 | 6 | 6 | 7 | 7 | 8 | 10 |
| Author1-non-Journal1 | | | | | | | | | | | 1 | 2 | 4 | 4 | 6 | 6 | 7 | 7 | 8 | 8 |
| | | | | | | | | | | | | | | | | | | | | |
| Author2-all | | | | | | | | | | | | | | 1 | 2 | 3 | 4 | 4 | 7 | 11 |
| Author2-non-Journal1 | | | | | | | | | | | | | | 1 | 2 | 2 | 4 | 4 | 6 | 7 |
| | | | | | | | | | | | | | | | | | | | | |
| Author3-all | | | | | | | | | | | | | | 1 | 2 | 2 | 2 | 2 | 2 | 4 |
| Author3-non-Journal1 | | | | | | | | | | | | | | 1 | 2 | 2 | 2 | 2 | 2 | 2 |
| Author4-all | 14 | 14 | 14 | 14 | 14 | 14 | 14 | 15 | 15 | 15 | 15 | 15 | 15 | 15 | 15 | 16 | 17 | 17 | 18 | 18 |
| Author4-non-Journal1 | 14 | 14 | 14 | 14 | 14 | 14 | 14 | 15 | 15 | 15 | 15 | 15 | 15 | 15 | 15 | 15 | 17 | 17 | 17 | 17 |
| Author5-all | 2 | 2 | 2 | 2 | 2 | 2 | 2 | 2 | 2 | 2 | 2 | 2 | 3 | 3 | 3 | 3 | 4 | 5 | 8 | 10 |
| Author5-non-Journal1 | 2 | 2 | 2 | 2 | 2 | 2 | 2 | 2 | 2 | 2 | 2 | 2 | 2 | 3 | 3 | 3 | 4 | 5 | 8 | 10 |

**Note:** Additions induced by Journal1's citations are marked via shading cells.

Author1 received 310 citations in about 10 years. The majority of citations occurred after 2013 (figure 2) and the increase was entirely due to citations within Journal1. Almost 95% (123 citations) of these Author1's citations were made after he/she had been appointed as an editor to Journal1. His/her recent citation pattern seems to increase in an "explosive" manner.

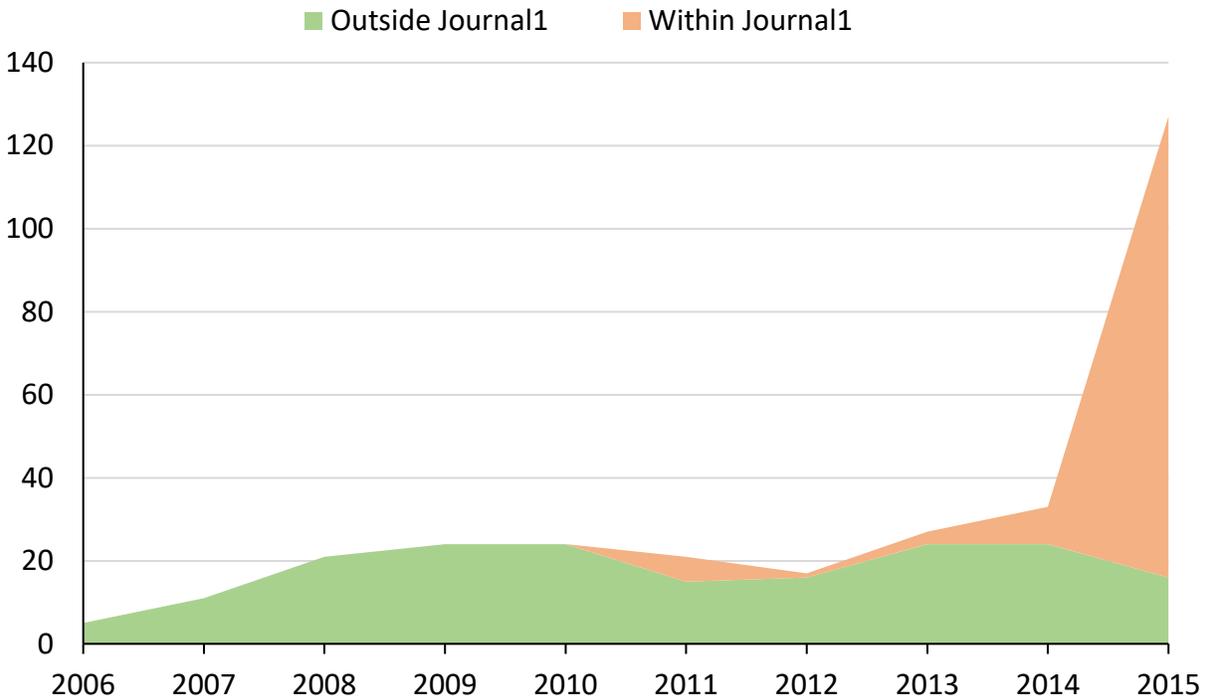

**Figure 2.** Citations of Author1's papers in WoS distinguishing between citations of Author1 papers, wherever published, whether cited or not in Journal1

## 2. Author2

Author2 became editor of Journal1 in July 2011 and associate editor to the same journal in January 2013. From WoS, Author2 has (co-)authored 55 papers (17 prior to his/her appointment to Journal1). He/she published 15 papers with Journal1 (only two prior to July 2011). His/her h-index is now 11 compared to three in 2011. If Journal1 is excluded as provider of citations his/her current *h*-index would be seven as compared with two in 2011 (table 3). As can be seen in figures 1 and 2, his/her works' citations interestingly follow the same pattern as that of Author1.

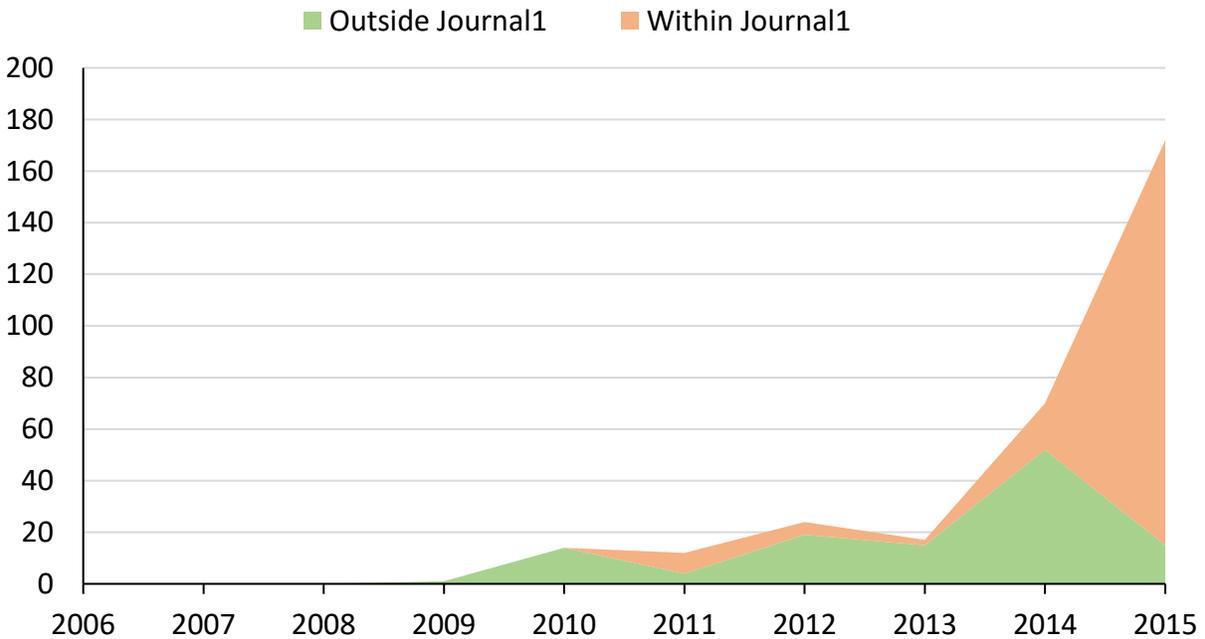

**Figure 3.** Citations of Author2's papers in WoS distinguishing between citations of Author2 papers, wherever published, whether cited or not in Journal1

Thus, both of them, after appointment to the editorial board of Journal1, have benefitted of citations in papers published in this same journal. That situation increased much in 2014 and became almost "explosive" in 2015 (figure 3).

Author2 total citations record amounts to 303 from which 183 (60%) are within Journal1; 40 are self-citations. Prior to 2012 there were eight citations received through Journal1 (including his/her own publications and papers authored by others). This means that the balance of 175 citations (more than 95%) occur after his/her appointment as a Journal1 editor.

Since 2008, Author1 and Author2 co-authored 11 papers that were cited 67 times in Journal1. Sometimes other co-authors also benefited from this situation, as will be seen.

### 3. Author3

Besides co-authorship activities between editors, as shown above, we must also consider the influence of editorial glorification [35] by citations by a co-author who is not an editor: Author3.

According to WoS, Author3 co-authored one paper each year in 2014 and 2015 with both Author1 and Author2 and one paper with Author1 in 2012. The research output of Author3 demonstrates 28 citations in Journal1 (19 of them being in a special issue released in October 2015). Between 2009 and 2013 his/her citations record is 18. Excluding papers published in Journal1 (no matter who authored them) in 2014 and 2015 his/her works received five and four citations (figure 4). His/her citation evolution in time follows the same pattern as for Author1 and Author2 (including the dependence by Journal1) but to a smaller scale.

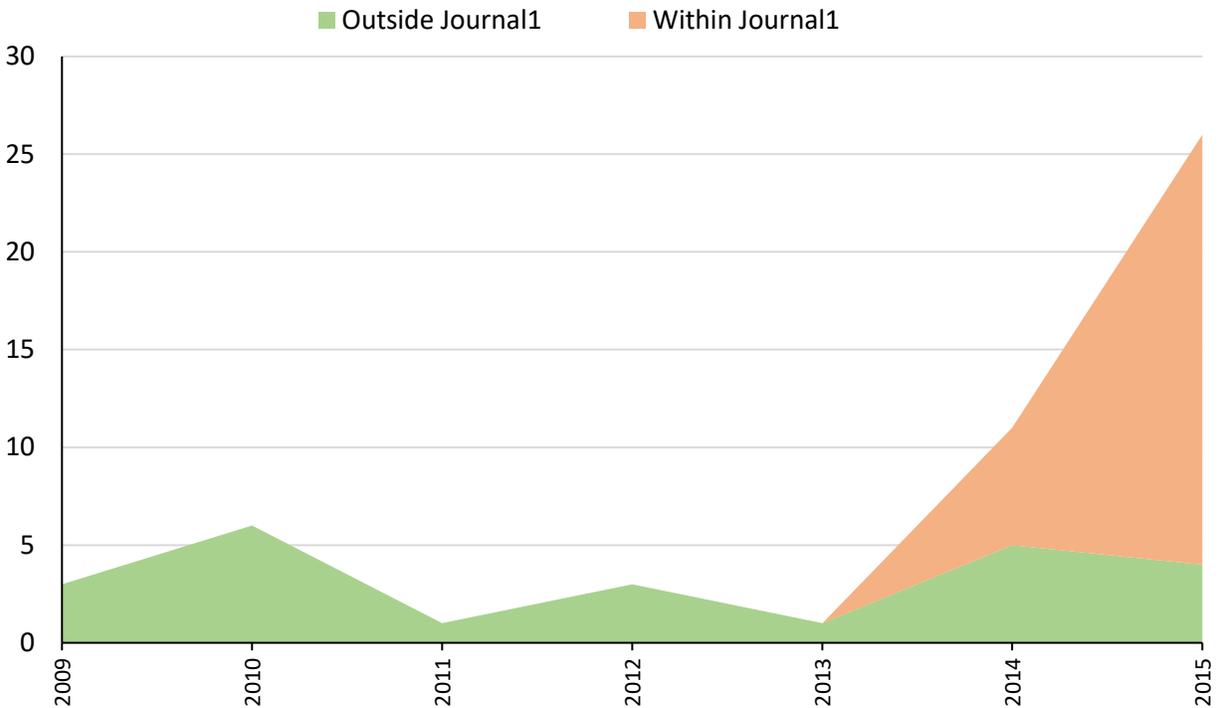

**Figure 4.** Citations of Author3's papers in WoS distinguishing between citations of Author3 papers, wherever published, whether cited or not in Journal1

In addition, due to his/her co-authorship (with Author1 and Author2), he/she received another six citations in 2014 and 22 in 2015 in Journal1 (regardless of who authored them). In so doing, his/her *h*-index doubled (from two to four) in 2015 (table 3). Thus, there is no doubt that the increase in Author3's prestige was a side-effect of his/her co-authorship to Author1 and Author2.

### 4. Author 4

As part of a control group, a similar set of tests as these used above was performed on the Editor in Chief (EIC), he/she has served as EIC of Journal1 for more than 25 years. According to WoS,

Author4 (co-)authored 196 manuscripts (in cooperation with 400 coauthors), including proceeding papers (96); papers (40) and book chapters (60). Author4 co-authored 4 papers published in Journal1 (all of these being published after his/her appointment as EIC). 77 manuscripts (co-)authored by Author4 were cited (between 1996 and 2015) at least once within WoS (A4ML).

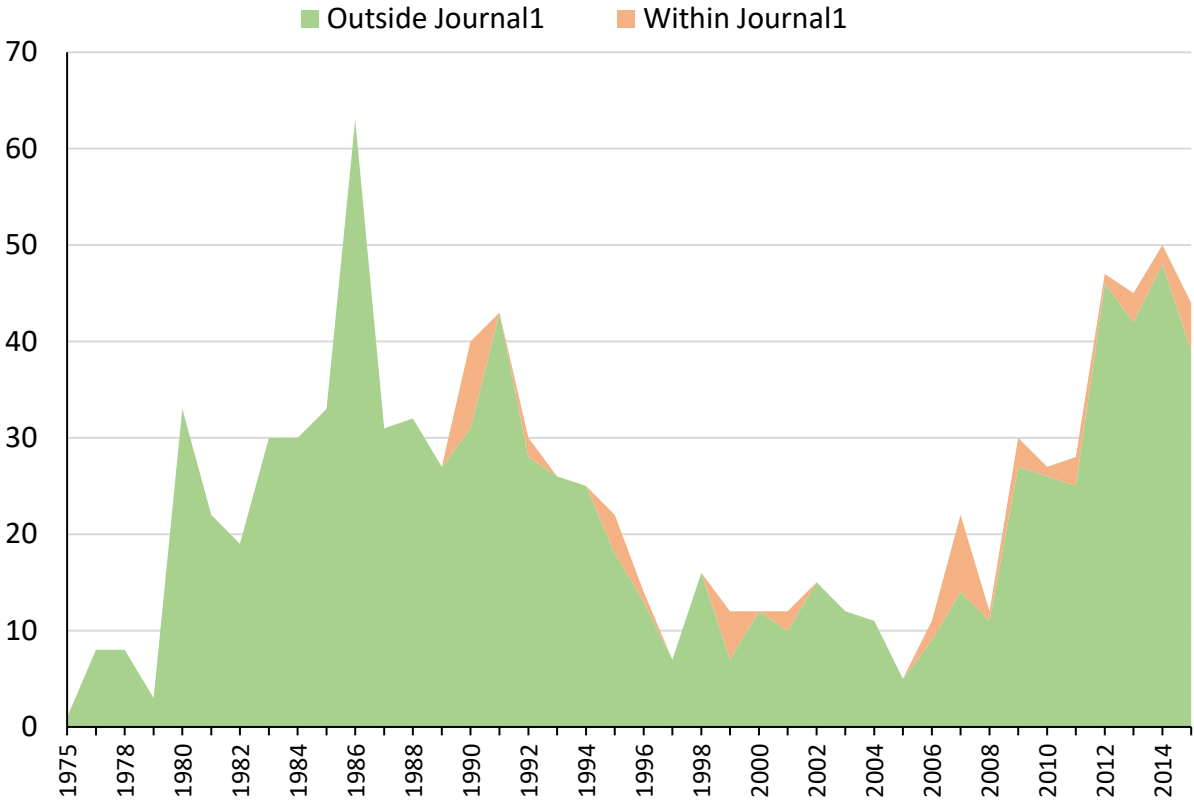

**Figure 5.** Citations of Author4's papers in WoS distinguishing between citations of Author4 papers, wherever published, whether cited or not in Journal1

Regarding citations (Figure 5), 958 citations (through the end of 2015) were found for papers (co-)authored by Author4. Even if Author4 serves as EIC, it was found that only 48 citations (a little bit more than 5% from the total 958 citations) in Journal1 (for papers published by him/her no matter in which venue) occurred. Author4's current (2015) *h*-index is 18, slightly smaller (by one unit) when not counting these (table 3). It is clear that the pattern of citations within Journal1 for Author4 is shows no dependence by Journal1.

## 5. Author 5

Another person who was appointed as an editor of Journal1 in the same timeframe as Author1 has been selected to be part of the control group: Author5. According to WoS, Author5 (co-)authored 47 manuscripts, most (40) of these being papers published in academic journals. Author5 co-authored three papers published within Journal1 (two of them being published prior to his/her appointment as editor of Journal1). An amount of 39 manuscripts co-authored by Author5 has been cited at least once (A5ML).

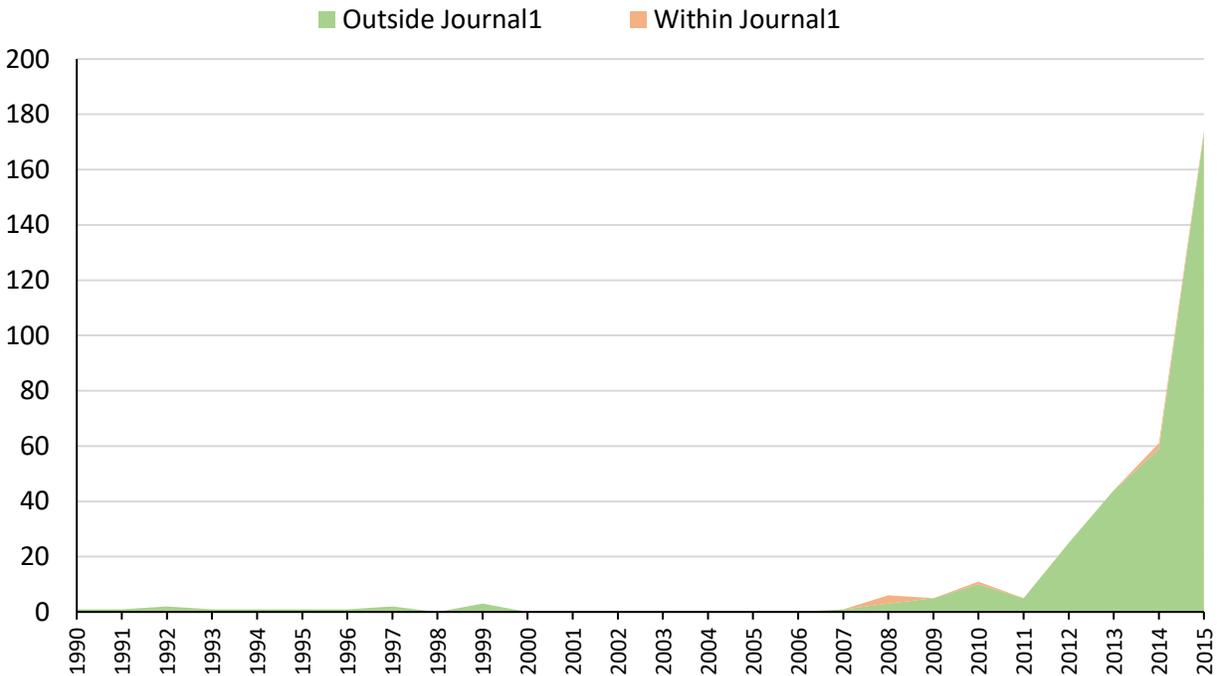

**Figure 6.** Citations of Author5's papers in WoS distinguishing between citations of Author5 papers, wherever published, whether cited or not in Journal1

From the perspective of his/her citations (Figure 6), although in the last few years his/her record grew rapidly, the relationship to Journal1 was neutral since only seven citations (out of 345) occurred in it. The share of citations from Journal1 for Author5 is only 2%. This neutrality is obvious because the portion of citations in Journal1 in Figure 6 is almost invisible. The evolution of the *h*-index (table 3) for Author5 is almost independent from the citations from Journal1. Thus, both persons (A4&A5) in the control group show neutrality in regards to Journal1's contribution to their citations records, which enables a favorable argument in support of the findings about the three other author/editors.

## 6. Where are citations coming from?

Let us show next that, as in previous inquiries [36], the Journal1 peer review process for regular issues seems to lead to less suggested additional references than does the process for special issues. Indeed, for papers published in regular issues of Journal1 (Table 4), more than 90% of coercive citations (as being signaled by our methodology), in regards to Author1, include only one "suggested" item. In the case of Journal1 special issues, in regards to the same author, almost two thirds of citations requested as additional references concern at least two papers.

**Table 4.** Number of Author1 paper citations according to journal's issue type

| Special Issue? | Number of Author1's items within the references list | | | Total |
|---|---|---|---|---|
| | 1 | 2 | 3 or more | |
| No | 28 (93.4%) | 1 (3.3%) | 1 (3.3%) | 30 (100%) |
| Yes | 12 (34.3%) | 13 (37.1%) | 10 (28.6%) | 35 (100%) |
| Total | 40 (61.5%) | 14 (21.5%) | 11 (17.0%) | 65 (100%) |

**Note:** Self and semi-self-citations were excluded.
The percentages are computed for each row.
The grouping factor is statistically significant (computed Chi Square test is 23.8056 while p value for 2 degrees of freedom is less than 0.01).

It can be concluded that the factor (special issue) provides statistically significant differences to consider them as non-respecting a null hypothesis: indeed, the p value for the Chi Square test is less than 0.01, as shown in Table 4. We are aware, of course, that these differences could occur also because: (i) the editor of the special issue is a highly recognized scholar in his/her field; subsequently he/she will get more citations; and (ii) the editor of the special issue has a good publication record on the topic and therefore bigger chances to get more citations.

**Table 5.** Number of Author2 paper citations according to journal's issue type

| Special Issue? | Number of Author2's items within the references list | | | Total |
|---|---|---|---|---|
| | 1 | 2 | 3 or more | |
| No | 35 (59.4%) | 13 (22.0%) | 11 (18.6%) | 59 (100%) |
| Yes | 28 (73.7%) | 6 (15.8%) | 4 (10.5%) | 38 (100%) |

| Total | 63 (65%) | 19 (19.6%) | 15 (15.5%) | 97 (100%) |

**Note:** Self and semi-self-citations were excluded.

The percentages are computed for each row.

The grouping factor is not statistically significant (computed Chi Square test is 3.7072 while p value for 2 degrees of freedom is 0.157).

As can be noticed in Table 5, Author2 has 97 distinct papers in Journal1, which leads to 183 citations (almost 1.9 cited Author2's items per paper in Journal1). However, a Chi Square test on the factor does not in this case lead to statistically significant values.

This approach was performed only for Author1 and Author2. Since the persons constituting the control group (Author4 and Author5) recorded only very small numbers of citations in Journal1 such deeper analysis is not deemed necessary nor especially useful since it neither supports nor undermines our points.

## Discussion and conclusions

We would like to point out that the unusual situation for the recognition of outstanding research was the trigger for our investigation of peer review patterns and brought about the current study. This paper is in line with detailed analyses of "explosive" author citations and the accompanying scientific glorification through manipulation of the *h*-index [13-14, 37-38].

It hardly needs to be said that it is quite normal, on the one hand, that special issues be published, sometimes through guest editors, and on the other hand that editorial boards evolve - in particular if some organizer has shown editorial skill. Notice that the promotion, addition, or removal, of editors could be of interest within the scope of our investigations, but gathering and analyzing such data would carry us far from the present aim, which concerns post-editorial board behavior.

Let us admit that the peer review process has always been criticized, mainly for its anonymous constraint. Advantages and defects cannot be easily quantified. Self-citations are also often frowned upon and have been discussed as a troubling parameter when using the *h*-index as a measure of work reputation. However, coercive citations are rarely discussed as a source of prestige. We show that it is a significant source. Of course, better access to reviewer reports (which actually happens for some journals like those published by Biomed Central or PeerJ) would be useful in order to measure the frequency and effect in scientific publishing fields.

Here, we have detected and demonstrated that one of the possible methods for increasing an author *h*-index is through activity on the editorial board of a scientific journal and filtering the papers for special issues, sometimes suggesting ad hoc references. A topic for further research concerns the affiliations of authors publishing in "special issues."

There are, indeed, other confounding variables (scientific domain, age, gender, geographic location, affiliation, content itself, journal visibility/rank of the cited paper, number of co-authors, productivity, career length etc.) that may affect the evolution of citations for an author. Only one journal is considered, which limits the academic domains involved in the study. While one could object about our "very small sample size," we maintain that this is not a significant impediment because it is accepted in academia [39] that a sample as small as a single case is sometimes sufficient to indicate the existence of a specific phenomenon or event. In fact, the sample size can be reasonably considered to be large enough and sufficiently varied to prove the thesis of the paper.

One may claim that the above data could be the result of the paper topics and spontaneous field and author recognition. The extremely fast increase in the number of citations for both "new" editors (Author1 and Author2) seems to show that this is not the case. Indeed, as pointed out above, the topic of the special issue could also be a trigger. Still there were plenty of citations which came through the papers published in regular issues.

Serving as an editor of an academic journal, in fact, appears to be a powerful position able to orient research and dissemination of knowledge. It also has the ability to influence the *h*-index of authors.

Of course, we do not claim that all citations are biased or coercive in a review process. It seems quite acceptable that reviewers and editors insist on pertinent references. The policy question remains whether editors should accept reviews suggesting an increase in references pertinent to the reviewers. In this respect, the role of editors is hard to analyze due to the lack of available data that is not usually released by editors and publishers (with rare exceptions). (We acknowledge comments by various editors emphasizing the time spent to control such apparently coercive citations.) We have shown in the present data analysis that a quantitative approach can be utilized to analyze data on editors without any breach of professional responsibility or etiquette.

Finally, recall that some hint that a journal could have editorial misconduct is found from the journal self-citation rate [40-41]. For example, JCR 2015 [34] shows that Journal1 has a 33% self-citation rate. Such a high self-citation rate is already an important signal for WoS, since journals that pass over the 50% threshold received a one year suspension from JCR [42].

The response rate of communication to the correspondent authors is small. It is, however, close to the standard for e-mail surveys [43]. This can be considered a limitation of the current study within the methodology of random surveys [44]. Replications of the current study are needed in order to confirm (or disconfirm) that coercive citations do exist. We recognize that our conclusions present challenges in the scientometrics field.

Further inquiries about this somewhat unethical approach of coercive citations should be done. Perhaps a way of avoiding coercive citations could be a more open peer review. Publishers (including big ones) should be aware of the existence of such situations and adapt their procedures accordingly.

## Acknowledgments

We approached corresponding authors who cited Author1's works. We thank all authors concerned who responded to questions and allowed us to substantiate the findings. Communication with the editor in chief of Journal1 should be acknowledged. Due to the electronic communication process Author1 approached the research team by email (twice) and phone (twice). Great thanks for Michael S. Jones who enhanced the English of an earlier version of this current manuscript. Thanks for Roxana Herteliu-Iftode who checked the English for the later version of our manuscript. MA work is part of scientific activities in the COST Action TD1306 "New Frontiers of Peer Review (PEERE)". CH, GR and TA work is part of scientific activities in the COST Action TD1210 "Analyzing the dynamics of information and knowledge landscapes (KNOWeSCAPE)". Highlights of this manuscript have been presented in the 2017 annual meeting of TD1210 COST Action in Sofia. We acknowledge important feedbacks received from the participants to this meeting.

## Author contribution

CH and BVI obtained data from WoS. MA, CH, BVI, GR and TA performed data analysis and manuscript design. Communication to papers' corresponding authors, Journal1 editor in chief and in-voluntarily to Author1 was responsibility of CH. CH and MA equally can be considered as being corresponding authors for this work.

**Competing interests:** CH, MA, BVI, GR and TA have nothing to declare regarding competing interests.